\documentclass[11pt,twoside]{article}

\usepackage{asp2004}
\usepackage{epsf}
\usepackage{lscape}
\usepackage{graphicx}

\begin{document}
\title{SPH simulations of accretion flow via Roche lobe overflow and via mass transfer from Be disk}
 \author{Kimitake~Hayasaki$^{1,3}$,
   Atsuo~T.~Okazaki$^2$ and James. R. Murray$^3$}
\affil{
$^1$Department of Applied Physics, Graduate School of Engineering,
        Hokkaido University, Kitaku N13W8, Sapporo 060-8628, Japan.\\
$^2$Faculty of Engineering, Hokkai-Gakuen University, Toyohira-ku,
      Sapporo 062-8605, Japan.\\
$^3$Centre for Astrophysics and Supercomputing, Swinburne University of Technology,
        Hawthorn Victoria 3122 Australia.}

\begin{abstract}
We compare the accretion flow onto the neutron star induced by Roche
lobe overflow with that by the overflow from the Be disk, in a zero
eccentricity, short period binary with the same mass transfer rate,
performing three-dimensional Smoothed Particle Hydrodynamics simulations.
We find that a persistent accretion disk is formed around the neutron
star in both cases.
The circularization radius of the material transferred via Roche lobe overflow is larger
than that of the material transfered from the Be disk.
Thus, the growth of the accretion disk in the former case becomes
significantly slower than in the latter case.
In both cases, the mass accretion rate is very small and varies little
with orbital phase, which is consistent with the observed X-ray
behaviour of Be/X-ray binaries with circular orbits (e.g. XTE
J1543-568).
\end{abstract}


\vspace{-0.5cm}
\section{Introduction} 

Most of binaries, which exhibit the X-ray activity, 
have a circular orbit around the common center of mass, 
in which an accretion disk is mainly formed via the
Roche lobe overflow from a mass-donor star.
On the other hand, in Be/X-ray binaries 
which consist of a neutron star and 
a Be star, 
an accretion disk is formed around the neutron star
via the mass transfer from the circumsteller disk of the Be star 
(Hayasaki \& Okazaki 2004).
Little work has been so far done 
on the accretion flow around the neutron star
with an attention to the difference 
between the overflow from the circumsteller disk around the Be star (model~A)
and the Roche lobe overflow from the Be star (model~B). 
In this paper, 
we study how the material accretes onto the neutron star
in both cases, performing three dimentional (3D) Smoothed 
Particle Hydrodynamics (SPH) simulations.

\section{Formation and Evolution of the accretion disk}

We carried out the simulations by using
the same 3D SPH code as 
in Hayasaki \& Okazaki (2004, 2005a and 2005b)
[see also Okazaki et al. (2002); Bate, Bonnell \& Price (1995)].
In order to investigate the accretion flow
around the neutron star,
two simulations have the same parameters as Hayasaki \& Okazaki (2005b)
except for the eccentricity $e=0.0$, the mass transfer rate 
$\dot{M}_{T}\sim1.24\times10^{-11}M_{\odot}yr^{-1}$
and the inner boundary radius $r_{\rm{in}}=6.0\times10^{-3}a$,
where $a$ is the semi-major axis of the binary.
In model~B, the Roche lobe overflow is modelled by launching the gas particles 
at the inner Laglange point $L_{1}$, in which  
we add $1000$ SPH particles per orbit
 with an initial speed $v_{inj}=0.1a\Omega_{\rm{orb}}$, in a direction $0.387$ rad prograde
of the binary axis (Lubow \& Shu 1975). 
In what follows, the units of time is $P_{\rm{orb}}=24.3\,d$. 

Fig.1 shows the orbital-phase dependence of the circularization radius (the left panel)
and the viscous time-scale in units of $P_{\rm{orb}}$ (the right panel).
In each panel, the solid line and the dotted line denote the circularization radius
and the viscous time-scale of model~A and model~B, respectively.
As shown in Fig.1, it is likely that the disk growth of model~A is 
faster than that of model~B.

Fig.2 gives snapshots of the accretion flow around the neutron star 
at $t=35$. Each panel shows the logarithm of the surface density, where
the solid curve denote the inner Roche lobe.
As seen in Fig.2, we note that a persistent accretion disk is formed around the neutron star in
both cases.

Fig.3 shows the evolution of mass-accretion rate.
The solid line and the dotted line show the 
mass-accretion rate of model~A and model~B, respectively.
We note from the figure that the mass accretion rate of model~A is much higher than 
that of model~B.

\begin{figure}[!ht]
\begin{center}
\includegraphics*[height=4.0cm]{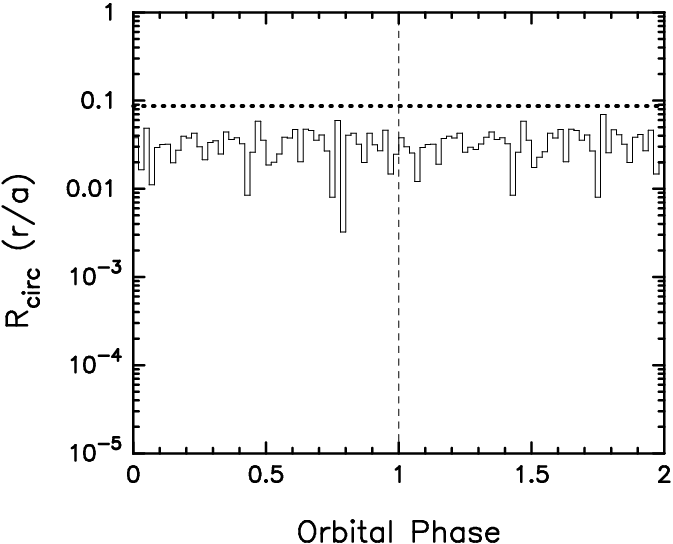}
\includegraphics*[height=4.0cm,angle=0]{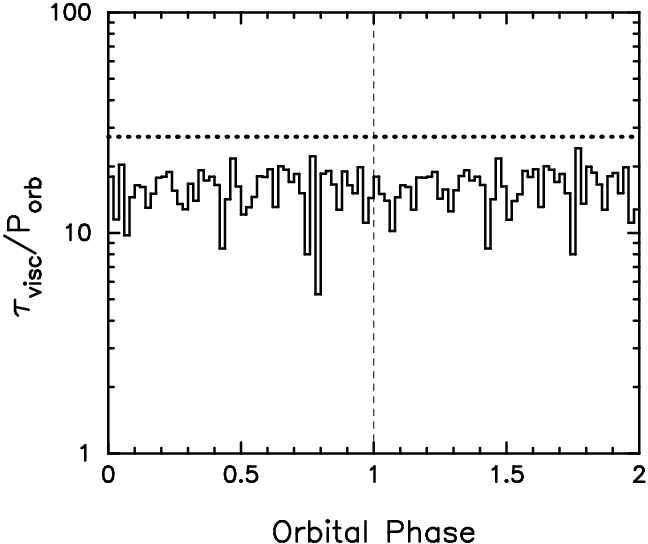}
\end{center}
\caption{
Orbital dependence of the circularization radius (the left panel) 
and the ratio of the viscous time-scale to the orbital period 
(the right panel). 
The solid line and the dotted line show the circularization radius 
and the viscous time-scale of the model~A and model~B, respectively.
}
\label{fig1}
\end{figure}


\begin{figure}[!ht]
\begin{center}
\includegraphics*[height=4.0cm,angle=270.0]{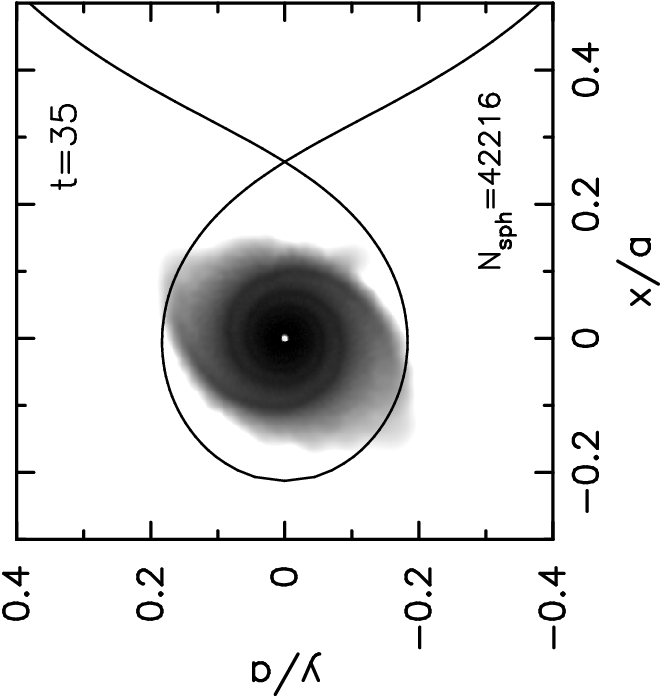}
\includegraphics*[height=4.0cm,angle=270.0]{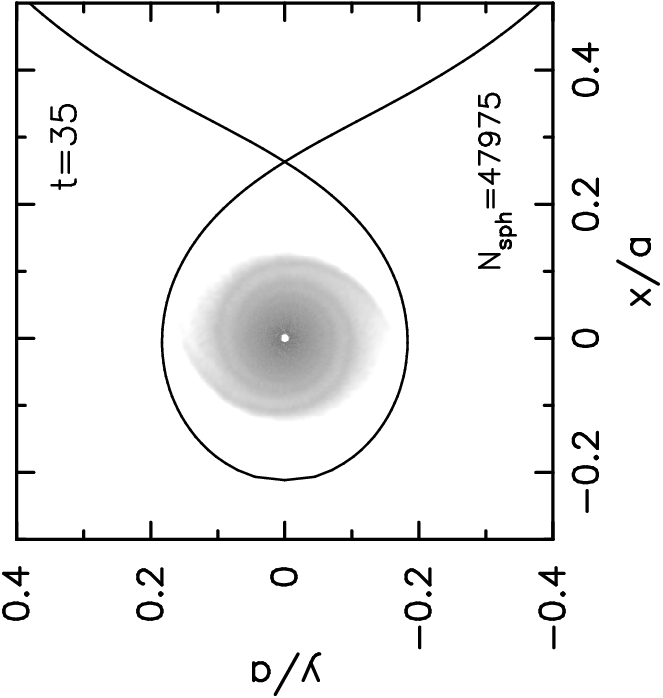}
\end{center}
\caption{
Snapshots of the accretion disk formation in a Be/X-ray binary
with $P_{\rm{orb}}=23.4$ and $e=0.0$ at $t=35.0$ 
in model~A (the left panel) and model~B (the right panel), 
respectively.
Each panel shows the surface density in a range of three orders of
magnitude in the logarithmic scale. Annotated in each panel are the
simulation time and the number of SPH particles.
The solid curve denote the inner Roche lobe in each panel.
}
\label{fig2}
\end{figure}


\begin{figure}[!ht]
\resizebox{\hsize}{!}
{\includegraphics{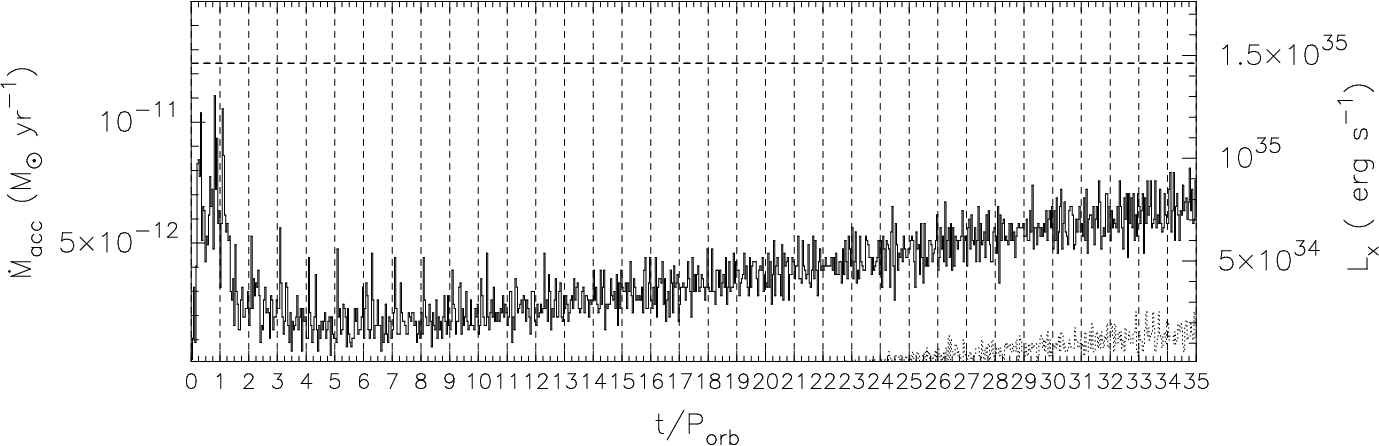}}
\caption{
Evolution of the mass-accretion rate for $0\le{t}\le35$.
The solid and dotted lines denote
the mass-accretion rate of model~A and model~B, respectively.
The dashed line shows the mean mass-transfer rate 
$\sim1.24\times10^{-11}~M_{\odot}/yr$ from the Be star.
The right axis denotes the X-ray luminocity corresponding 
to the mass-accretion rate.
}
\label{fig3}
\end{figure}

\vspace{-1.0cm}
\section{Summary}
We have perfomed 3D SPH simulations in order to
compare the accretion flow onto the neutron star induced by 
Roche lobe overflow with that by the overflow from the Be disk in a
a zero eccentricity, short period binary with the same mass-transfer rate.
We have found that a persistent accretion disk is formed around the neutron star in 
both cases, as seen in Fig.2.
The mass accretion rate in model~A is much higher than that 
in model~B becasue of the smaller circularization radius in model~A, 
as shown in Fig.1 and 3. 
This indicates that the disk growth is faster
as the specific angular momentum of gas particles is lower.
In either case, the tiny mass-accretion rate 
is consistent with the
observed X-ray
behaviour of Be/X-ray binaries with circular orbits (e.g. XTE
J1543-568).

\acknowledgements The simulations reported here were performed using the facility
of the Centre for Astrophysics \& Supercomputing at
Swinbune University of Technology, Australia.
This work has been supported by Grant-in-Aid for the 21st Century
COE Scientific Research Programme on "Topological Science and Technology"
from the Ministry of
Education, Culture, Sport, Science and Technology of Japan (MECSST) and
in part by Grant-in-Aid for Scientific Reserch
(15204010) of Japan Society for the Promotion of Science.


\end{document}